\documentclass[11pt]{article}
\usepackage[utf8]{inputenc}
\usepackage[T1]{fontenc}
\usepackage{authblk}
\usepackage{amsmath,amssymb,amsfonts,amsthm}
\usepackage[top=.80in, bottom=.80in, left=.80in, right=.80in]{geometry}
\newcommand{\Keywords}[1]{\par\noindent
{\small{\em Keywords\/}: #1}}

\catcode`,\active

\catcode`\,12

\newcommand*\pFqskip{8mu}
\catcode`,\active
\newcommand*\pFq{\begingroup
        \catcode`\,\active
        \def ,{\mskip\pFqskip\relax}%
        \dopFq
}
\catcode`\,12
\def\dopFq#1#2#3#4#5{%
        {}_{#1}F_{#2}\biggl[\genfrac..{0pt}{}{#3}{#4};#5\biggr]%
        \endgroup
}

\newcommand{\unit}{1\!\!1}
\newcommand{\class}[1]{\par\noindent
{\small{AMS classification scheme numbers\/}: #1}}
\newcommand{\pd}{\partial}
\newcommand{\ket}[1]{|#1\rangle}

\newcommand{\braket}[2]{\langle #1|#2\rangle}

\newcommand{\under}[1]{_{#1}}
\numberwithin{equation}{section}
\title{Superintegrability in two dimensions\\ and the Racah-Wilson algebra}
\date{}
\author[1]{Vincent X. Genest\thanks{genestvi@crm.umontreal.ca}}
\author[1]{Luc Vinet\thanks{luc.vinet@umontreal.ca}}
\author[2]{Alexei Zhedanov\thanks{zhedanov@kinetic.ac.donetsk.ua}}
\affil[1]{Centre de Recherches Math\'ematiques, Universit\'e de Montr\'eal, C.P. 6128, Succursale Centre-ville, Montr\'eal, Qu\'ebec, H3C 3J7, Canada}
\affil[2]{Donetsk Institute for Physics and Technology, Donetsk 83114, Ukraine}
\begin{document}
\maketitle
\thispagestyle{empty}
\hrule
\begin{abstract}
\noindent
The analysis of the most general second-order superintegrable system in two dimensions: the generic 3-parameter model on the 2-sphere, is cast in the framework of the Racah problem for the $\mathfrak{su}(1,1)$ algebra. The Hamiltonian of the 3-parameter system and the generators of its quadratic symmetry algebra  are seen to correspond to the total and intermediate Casimir operators of the combination of three $\mathfrak{su}(1,1)$ algebras, respectively. The construction makes explicit the isomorphism between the Racah-Wilson algebra, which is the fundamental algebraic structure behind the Racah problem for $\mathfrak{su}(1,1)$, and the invariance algebra of the generic 3-parameter system. It also provides an explanation for the occurrence of the Racah polynomials as overlap coefficients in this context. The irreducible representations of the Racah-Wilson algebra are reviewed as well as their connection with the Askey scheme of classical orthogonal polynomials.
\\

\\
\class{81R12,33C45}
\end{abstract}
\hrule
\section{Introduction}
The purpose of this paper is to stress the algebraic equivalence between the analysis of the generic quantum 3-parameter superintegrable system on the 2-sphere and the Racah problem of $\mathfrak{su}(1,1)$. The equivalence will be made explicit by the direct identification of the natural operators of the Racah problem, i.e. the intermediate Casimir operators for the combination of three $\mathfrak{su}(1,1)$ algebras, with the symmetry operators that span the invariance algebra of the system. From this identification will follow the explicit isomorphism between the symmetry algebra of the 3-parameter model and the Racah-Wilson algebra, which is the algebra behind the Racah problem for $\mathfrak{su}(1,1)$. Since the Racah-Wilson algebra is also the algebraic structure that encodes the properties of the Racah and Wilson polynomials, the isomorphism will also provide an elegant explanation for the occurrence of these families of polynomials as overlap coefficients in the 3-parameter system on the 2-sphere.

\subsection{The Racah-Wilson algebra}
The Racah-Wilson algebra is the infinite-dimensional associative algebra generated by the algebraically independent elements $K_1$, $K_2$ that satisfy, together with their commutator $K_3\equiv [K_1,K_2]$, the following commutation relations:
\begin{align}
\begin{aligned}\label{Racah-Wilson}
[K_2,K_3]&=a_2 K_2^2+a_1\{K_1,K_2\}+c_1 K_1+d K_2+e_1,
\\
[K_3,K_1]&=a_1 K_1^2+a_2 \{K_1,K_2\}+c_2 K_2+dK_1+e_2,
\end{aligned}
\end{align}
where the structure constants are assumed to be real (it will be seen that the number of structure constants can be reduced from seven to three by affine transformations of the generators). The algebra \eqref{Racah-Wilson} first appeared in the coupling  problem of three angular momenta, i.e. the Racah problem for $\mathfrak{su}(2)$ \cite{Zhedanov-1988}. It was first observed in \cite{Zhedanov-1988} that the intermediate Casimir operators in the combination of three angular momenta realize the algebra \eqref{Racah-Wilson} and this observation was exploited to derive the symmetry group of the $6j$-symbol (Racah coefficients). In the same paper, a special case of \eqref{Racah-Wilson} corresponding to $a_1=0$ was also seen to occur in the Clebsch-Gordan problem of $\mathfrak{su}(2)$ and the algebraic relations were used to find the symmetries of the $3j$ coefficients. The representations of the algebra \eqref{Racah-Wilson} were presented to some extent in \cite{Zhedanov-1989} and the link between these representations and the Askey scheme of classical orthogonal polynomials was established. More specifically, it was shown that for certain finite(infinite)-dimensional irreducible representations of \eqref{Racah-Wilson}, the Racah (Wilson) polynomials occur as overlap coefficients between the eigenbases of $K_1$ and $K_2$. Furthermore, a suitable generalization of \eqref{Racah-Wilson} whose representations encompass the full Askey scheme of basic or $q$-orthogonal polynomials was also proposed; this generalization is now referred to as the Askey-Wilson algebra \cite{Zhedanov-1991-1}. A number of papers followed in which special cases of the algebra \eqref{Racah-Wilson} were seen to occur as symmetry algebras of classical and quantum second-order superintegrable systems in two dimensions \cite{Zhedanov-1991-2,Zhedanov-1991-3,Zhedanov-1992-2}, some of them based on the combination of two $\mathfrak{su}(1,1)$ algebras \cite{Zhedanov-1993}, i.e. the Clebsch-Gordan problem for $\mathfrak{su}(1,1)$. Quite interestingly, the full Racah algebra was not encountered then. It was only later, with the complete classification of second-order superintegrable systems, that a significant step in this direction was made in the study of the generic 3-parameter system on the 2-sphere \cite{Kalnins-2007}. Before discussing this particular model and the results of \cite{Kalnins-2007}, let us first recall a few points about quantum superintegrability.
\subsection{Superintegrability}
A quantum system described by a Hamiltonian $H$ with $d$ degrees of freedom is \emph{maximally superintegrable} if it admits $2d-1$ algebraically independent symmetry operators $S_i$ satisfying
\begin{align*}
[H,S_i]=0,\quad 1\leqslant i \leqslant 2d-1,
\end{align*}
where one the symmetries is the Hamiltonian itself, e.g. $S_1\equiv H$. For such a system, it is impossible for all the $S_i$ to commute with one another and hence the symmetries generate a non-Abelian \emph{invariance algebra} for $H$. In practice, the associated Schr\"odinger equation
\begin{align*}
H\Psi=E\psi,
\end{align*}
can be exactly solved both analytically and algebraically. A superintegrable system is said to be order $\ell$ if $\ell$ is the maximum order of the symmetries (excluding $H$) in the momentum variables. First order superintegrable systems ($\ell=1$) have geometrical symmetries and their invariance algebras are Lie algebras \cite{Olver-1993}. Second order superintegrable systems ($\ell=2$) typically admit separation of variables in more than one coordinate system and have quadratic symmetry algebras \cite{Zhedanov-1992,Granovskii-1992,Vinet-1995,Winter-2005}. In two dimensions, all first and second order superintegrable systems are known and have been classified \cite{Daska-2006,Winter-1965,Kalnins-2000,Kalnins-2000-2,Winter-1967}. The study of superintegrable systems is important in its own right in view of their numerous applications. They also constitute a bedrock for the analysis of symmetries.

\subsection{The generic 3-parameter system on the 2-sphere}
In two dimensions, one of the most important second-order superintegrable systems is the generic 3-parameter model on the 2-sphere \cite{Miller-2013}. This model is described by the Hamiltonian
\begin{align}
\label{Hamiltonian}
\mathcal{H}=J_1^2+J_2^2+J_3^2+\frac{k_1^2-\frac{1}{4}}{x_1^2}+\frac{k_2^2-\frac{1}{4}}{x_2^2}+\frac{k_3^2-\frac{1}{4}}{x_3^2},
\end{align}
with the constraint $x_1^2+x_2^2+x_3^2=1$. The operators $J_{i}$, $i=1,2,3$, stand for the standard angular momentum generators
\begin{align}
\label{Def-J}
J_1=-i(x_2\pd_{x_3}-x_3\pd_{x_2}),\quad J_2=-i(x_3\pd_{x_1}-x_1\pd_{x_3}),\quad J_3=-i(x_1\pd_{x_2}-x_2\pd_{x_1}),
\end{align}
which satisfy the familiar $\mathfrak{so}(3)$ commutation relations
\begin{align}
[J_i,J_j]=i\sum_{k=1}^{3}\epsilon_{ijk}J_{k},
\end{align}
where $\epsilon_{ijk}$ is the totally antisymmetric tensor and $\pd_{x_i}$ denotes differentiation with respect to the variable $x_i$. It is known (see \cite{Miller-2013} and references therein) that all second-order superintegrable models in two dimensions are limiting cases of \eqref{Hamiltonian}; hence the generic 3-parameter model on the 2-sphere described by the Hamiltonian \eqref{Hamiltonian} can be considered as the most general system of such type. The symmetries of $\mathcal{H}$ and the quadratic invariance algebra they span can be found in \cite{Kalnins-1996,Kalnins-2000-2}. Upon taking
\begin{align}
\label{Def-Symmetries}
L_1=J_1^2+\frac{a_2 x_3^2}{x_2^2}+\frac{a_3 x_2^2}{x_3^2},\quad L_2=J_2^2+\frac{a_3 x_1^2}{x_3^2}+\frac{a_1 x_3^2}{x_1^2},\quad L_3=J_3^2+\frac{a_1 x_2^2}{x_1^2}+\frac{a_2 x_1^2}{x_2^2},
\end{align}
where $a_1=k_1^2-1/4$, $a_2=k_2^2-1/4$ and $a_3=k_3^2-1/4$, it is directly checked that $[\mathcal{H},L_i]=0$, i.e. that the operators $L_i$ are constants of motion. Furthermore, upon defining $R=[L_1,L_2]$, the following commutation relations hold:
\begin{align}
\label{Symmetry-Algebra-1}
[L_i,R]=4\{L_i,L_j\}-4\{L_i,L_k\}-(8-16a_j)L_j+(8-16a_k)L_{k}+8(a_j-a_k),
\end{align}
where $\{x,y\}=xy+yx$ stands for the anticommutator. The four symmetry operators $L_1$, $L_2$, $L_3$, $R$ and the Hamiltonian $\mathcal{H}$ are not algebraically independent from one another. Indeed, one has 
\begin{align}
\label{H-J}                                                                                                                                                                       \mathcal{H}=L_1+L_2+L_3+a_1+a_2+a_3,                                                                                                                                                               \end{align}
which relates the sum of the three symmetries \eqref{Def-Symmetries} to the Hamiltonian. Moreover, the square of the symmetry operator $R$ can be shown to satisfy \cite{Kalnins-2000-2}
\begin{align}
\begin{aligned}\label{Symmetry-Algebra-2}
R^2=&-\frac{8}{3}\,\{L_1,L_2,L_3\}-\sum_{i=1}^{3}\left\{(12-16a_i)L_i^2+\frac{1}{3}(16-176a_i)L_i+\frac{32}{3}a_i\right\}\\
&+\frac{52}{3}\left(\{L_1,L_2\}+\{L_2,L_3\}+\{L_1,L_3\}\right)+48(a_1a_2+a_2a_3+a_3a_1)-64\,a_1a_2a_3,
\end{aligned}
\end{align}
where $\{x_1,x_2,x_3\}$ is the symmetrized sum of six terms of the form $x_ix_jx_k$. In view of the relations \eqref{H-J} and \eqref{Symmetry-Algebra-2}, it is clear that the system described by the Hamiltonian \eqref{Hamiltonian} possesses three algebraically independent symmetries: $\mathcal{H}$, $L_1$, $L_2$, and is hence maximally superintegrable. Moreover, since $L_1$, $L_2$ are second order differential operators, the generic 3-parameter system of the sphere is superintegrable of order $\ell=2$. The solutions of the Schr\"odinger equation associated to $\mathcal{H}$ have been obtained in \cite{Kalnins-1996} by separation of variables in various coordinate systems. In spherical coordinates, these solutions are given in terms of the classical Jacobi polynomials.
\subsection{The 3-parameter system and Racah polynomials}
In a remarkable paper \cite{Kalnins-2007}, the finite(infinite)-dimensional irreducible  representations of the symmetry algebra \eqref{Symmetry-Algebra-1}, \eqref{Symmetry-Algebra-2} have been related to the Racah (Wilson) polynomials. More specifically, it was shown that this symmetry algebra can be realized in terms of difference operators of which the Racah (Wilson) polynomials are eigenfunctions. Given the complexity of the algebraic relations \eqref{Symmetry-Algebra-1}, \eqref{Symmetry-Algebra-2}, the result constitutes a \emph{tour de force}. It also implies that the Racah polynomials act as overlap coefficients between the eigenbases of $L_1$ and $L_2$ in a given energy eigenspace of the Hamiltonian \eqref{Hamiltonian}. Moreover it allows, through contractions \cite{Miller-2013}, to tie the entire Askey scheme of classical orthogonal polynomials to physical systems. 

The occurrence of the Racah (Wilson) polynomials in the representations of the symmetry algebra of the 3-parameter model strongly suggests that the symmetry algebra \eqref{Symmetry-Algebra-1}, \eqref{Symmetry-Algebra-2} can be explicitly written in the form \eqref{Racah-Wilson}. More importantly, it indicates the existence of a connection between the 3-parameter system and the Racah problem of either $\mathfrak{su}(1,1)$ or $\mathfrak{su}(2)$, for which the Racah-Wilson algebra is the underlying algebraic structure. As suggested by the work of Kuznetsov \cite{Kuz-1992} and by \cite{Zhedanov-1993}, the connection will be made through the $\mathfrak{su}(1,1)$ algebra.  It will be shown explicitly that the analysis of the generic 3-parameter system on the 2-sphere is equivalent  to the Racah problem for $\mathfrak{su}(1,1)$. Using a realization of $\mathfrak{su}(1,1)$ in terms of differential operators, the 3-parameter Hamiltonian \eqref{Hamiltonian} will be seen to correspond to the total Casimir operator of the combination of three $\mathfrak{su}(1,1)$ algebras and the symmetries $L_1$, $L_2$, $L_3$ will be identified with the intermediate Casimir operators. From this identification will follow the explicit isomorphism between the invariance algebra \eqref{Symmetry-Algebra-1}, \eqref{Symmetry-Algebra-2} and the Racah-Wilson algebra which, as we recall, is the fundamental algebraic structure for both the Racah problem for $\mathfrak{su}(1,1)$ and the Askey-Scheme of classical orthogonal polynomials. 
\subsection{Outline}
The outline of the paper is as follows. In section 2, the connection between the finite-dimensional irreducible representations of Racah-Wilson algebra and the Racah polynomials is established. This connection is obtained in two equivalent ways. First, the finite-dimensional irreducible representations of the Racah-Wilson algebra are developed in a model-independent fashion and are then related to the Racah polynomials. Second, it is shown that the difference operators corresponding to the difference equation and recurrence relation of the Racah polynomials realize the Racah-Wilson algebra \eqref{Racah-Wilson}. The reader who wishes to focus on the connection between the Racah problem and superintegrability could proceeed directly to subsection 2.4. In section 3, the analysis of the Racah problem for $\mathfrak{su}(1,1)$ by means of the Racah algebra is revisited. In section 4, the equivalence between the Racah problem for $\mathfrak{su}(1,1)$ and the generic 3-parameter superintegrable system is established using a differential realization of $\mathfrak{su}(1,1)$. The isomorphism between the invariance algebra of the model and the quadratic Racah algebra is written down explicitly. Some perspectives on future investigations are offered in the conclusion.
\section{Representations of the Racah-Wilson algebra}
In this section, a review of the construction of the irreducible representations of the Racah-Wilson algebra is presented. Another presentation can be found in \cite{Zhedanov-1989}. Here the emphasis is put on the finite-dimensional representations and on the relation between these representations and the classical Racah polynomials.
\subsection{The Racah-Wilson algebra and its ladder property}
Recall that the defining relations  of the Racah-Wilson algebra have the expression
\begin{subequations}
\label{QR}
\begin{align}
[K_1,K_2]&=K_3,\\
[K_2,K_3]&=a_2 K_2^2+a_1 \{K_1,K_2\}+c_1 K_1+d K_2+e_1,\\
[K_3,K_1]&=a_1 K_1^2+a_2 \{K_1,K_2\}+c_2 K_2+d K_1+e_2,
\end{align}
\end{subequations}
where it is assumed that $a_1\cdot a_2\neq 0$. The algebra admits the Casimir operator \cite{Zhedanov-1988}
\begin{align}
\begin{aligned}
\label{QR(3)-Casimir}
Q=&a_1 \{K_1^2,K_2\}+a_2 \{K_1,K_2^2\}+K_3^2+(a_1^2+c_1)K_1^2+(a_2^2+c_2)K_2^2\\
&+(d+a_1a_2)\{K_1,K_2\}+(da_1+2e_1)K_1+(d a_2+2e_2)K_2,
\end{aligned}
\end{align}
which commutes with all generators. It can be seen that the number of parameters in \eqref{QR(3)} can be reduced from seven to three by taking the linear combinations $K_1\rightarrow u_1 K_1+v_1$, $K_2\rightarrow u_2 K_2+v_2$, $K_3\rightarrow u_1u_2K_3$ and adjusting the coefficients $u_i$, $v_i$. A convenient choice for the study of the representations of \eqref{QR(3)} is obtained by taking
\begin{align}
u_1=a_2^{-1},\quad u_2=a_1^{-1},\quad v_1=c_2/2a_2^2,\quad v_2=c_1/2a_1^2.
\end{align}
This leads to the following reduced form for the defining relations of the Racah-Wilson algebra:
\begin{subequations}
\label{QR(3)}
\begin{align}
\label{QR(3)-a}
[K_1,K_2]&=K_3,\\
\label{QR(3)-b}
[K_2,K_3]&=K_2^2+\{K_1,K_2\}+d K_2+e_1,\\
\label{QR(3)-c}
[K_3,K_1]&=K_1^2+\{K_1,K_2\}+d K_1+e_2,
\end{align}
\end{subequations}
which contains only three parameters $d$, $e_1$, $e_2$. The Casimir \eqref{QR(3)-Casimir} for the algebra \eqref{QR(3)} is of the form
\begin{align}
\begin{aligned}
\label{QR(3)-Casimir-2}
Q=&\{K_1^2,K_2\}+\{K_1,K_2^2\}+K_1^2+K_2^2+K_3^2
\\
&+(d+1)\{K_1,K_2\}+(2e_1+d)K_1+(2e_2+d)K_2,
\end{aligned}
\end{align}
One of the most important characteristics of the Racah-Wilson algebra is that it possesses a ``ladder'' property. To exhibit this property, let $\omega_{p}$ be an eigenvector of $K_1$ with eigenvalue $\lambda_{p}$
\begin{align}
K_1\omega_{p}=\lambda_{p}\omega_{p},
\end{align}
where $p$ is an arbitrary real parameter. One can construct a new eigenvector $\omega_{p'}$ corresponding to a different eigenvalue $\lambda_{p'}$ by taking
\begin{align}
\label{Rel-1}
\omega_{p'}=\left\{\alpha(p)K_1+\beta(p) K_2+\gamma(p) K_3\right\}\omega_{p},
\end{align}
where $\alpha(p)$, $\beta(p)$, $\gamma(p)$ are coefficients to be determined. Upon combining the eigenvalue equation for $\omega_{p'}$
\begin{align}
K_1\omega_{p'}=\lambda_{p'}\omega_{p'}
\end{align}
with \eqref{Rel-1}, using the commutation relations \eqref{QR(3)-a}, \eqref{QR(3)-c} and solving for the coefficients $\alpha(p)$, $\beta(p)$, $\gamma(p)$, it easily seen that the eigenvalues $\lambda_{p}$, $\lambda_{p'}$ must satisfy
\begin{align}
\label{Rel-2}
(\lambda_{p'}-\lambda_{p})^2+(\lambda_{p'}+\lambda_{p})=0.
\end{align}
For a given value of $\lambda_{p}$, the quadratic equation \eqref{Rel-2} yields two possible values for $\lambda_{p'}$, say $\lambda_+$ and $\lambda_-$. Without loss of generality, one can define $\lambda_+=\lambda_{p+1}$ and $\lambda_{-}=\lambda_{p-1}$. We assume the eigenvalues to be non-degenerate and denote by $E_{\lambda}$ the corresponding one-dimensional eigenspaces. It then follows from the analysis above that a generic ($p$-dependent) algebra elements maps $E_{\lambda_{p}}\rightarrow E_{\lambda_{p-1}}\oplus E_{\lambda_{p}}\oplus E_{\lambda_{p+1}}$. The element $K_2$ is thus 3-diagonal and since $K_3=K_1K_2-K_2K_1$ with $K_1$ diagonal, $K_3$ is 2-diagonal. In the basis with vectors $\omega_{p}$, one therefore has
\begin{align}
\begin{aligned}
\label{Tri-Diag}
K_1\,\omega_p&=\lambda_{p}\,\omega_{p},\\
K_2\,\omega_{p}&=A_{p+1}\,\omega_{p+1}+B_{p}\,\omega_{p}+A_{p}\,\omega_{p-1},\\
K_3\,\omega_{p}&=g_{p+1}A_{p+1}\,\omega_{p+1}-g_{p}A_{p}\,\omega_{p-1},
\end{aligned}
\end{align}
where $g_{p}=\lambda_{p}-\lambda_{p-1}$. Note that for $K_2$ to be self-adjoint $A_{p}$ has to be real. The result \eqref{Tri-Diag} will now be specialized to finite-dimensional irreducible representations. Note that the result \eqref{Tri-Diag} has been used in \cite{Zhedanov-1989} to derive infinite-dimensional representations for which the Wilson polynomials act as overlap coefficients between the respective eigenbases of the independent generators $K_1$, $K_2$.

\subsection{Discrete-spectrum and finite-dimensional representations}
In finite-dimensional irreducible representations, the spectrum of $K_1$ is discrete and thus one can denote the eigenvectors of $K_1$ by $\psi_{n}$ with $n$ an integer. Then by \eqref{Tri-Diag} one may write for the actions of the generators
\begin{subequations}
\label{Actions-2}
\begin{align}
K_1\psi_{n}&=\lambda_{n}\psi_{n},\\
\label{Actions-2-b}
K_2\psi_{n}&=A_{n+1}\psi_{n+1}+B_{n}\psi_{n}+A_{n}\psi_{n-1},\\
K_3\psi_{n}&=A_{n+1}g_{n+1}\psi_{n+1}-A_{n}g_{n}\psi_{n-1},
\end{align}
\end{subequations}
where $g_{n}=\lambda_n-\lambda_{n-1}$ and $A_n$ is real. Upon substituting the actions \eqref{Actions-2} in the commutation relation \eqref{QR(3)-c} and using \eqref{QR(3)-a}, one finds that the eigenvalues $\lambda_{n}$ satisfy
\begin{align}
\label{Equality-4}
(\lambda_{n+1}-\lambda_{n})^2+(\lambda_n+\lambda_{n+1})=0.
\end{align}
The recurrence relation \eqref{Equality-4} admits two solutions differing only by the sign of the integration constant. Without loss of generality, one can thus write
\begin{align}
\label{K1-Eigenvalues}
\lambda_{n}=-(n-\sigma)(n-\sigma+1)/2,
\end{align}
where $\sigma$ is a arbitrary real parameter. From \eqref{QR(3)-c} and the eigenvalues \eqref{K1-Eigenvalues}, one can evaluate $B_n$ and $g_n$ directly to find
\begin{align}
\label{Bn-Gn}
B_n=-\frac{\lambda_n^2+d\,\lambda_n+e_2}{2\lambda_{n}},\quad g_n=(\sigma-n).
\end{align}
At this stage, the actions \eqref{Actions-2} with \eqref{K1-Eigenvalues} and \eqref{Bn-Gn} are such that the relations \eqref{QR(3)-a} and \eqref{QR(3)-c} are satisfied. There remains only to evaluate $A_{n}$ in \eqref{Actions-2}. Upon acting with relation \eqref{QR(3)-b} on $\psi_{n}$, using \eqref{Actions-2} and \eqref{Bn-Gn} and then gathering the terms in $\psi_{n}$, one obtains the following recurrence relation for $A_{n}^2$:
\begin{align}
\label{Equality-6}
2\Big\{g_{n+3/2}\,A_{n+1}^2-g_{n-1/2}\,A_{n}^2\Big\}=B_{n}^2+(2\lambda_n+d)B_{n}+e_1.
\end{align}
Instead of solving the recurrence relation \eqref{Equality-6}, one can use the Casimir operator \eqref{QR(3)-Casimir} to obtain $A_{n}^2$ directly. One first requires that the Casimir operator acts as a multiple of the identity on the basis $\psi_{n}$, as is demanded by Schur's lemma for irreducible representations. Hence one takes
\begin{align}
\label{Equality-5}
Q\psi_{n}=q\,\psi_{n}.
\end{align}
Upon substituting \eqref{QR(3)-Casimir} in \eqref{Equality-5} with the actions \eqref{Actions-2} and then gathering the terms in $\psi_{n}$, a straightforward computation yields
\begin{align}
\begin{aligned}
\label{Equality-7}
2\Big\{g_{n+3/2}g_{n}A_{n+1}^2+g_{n+1}g_{n-1/2} A_{n}^2\Big\}&=(2\lambda_n+1)B_n^2+\lambda_n^2+(2e_1+d)\lambda_n-q\\
&+(2\lambda_n^2+2\lambda_n(d+1)+2e_2+d) B_n.
\end{aligned}
\end{align}
Combining \eqref{Equality-6} and \eqref{Equality-7}, one obtains
\begin{align}
\label{Equality-8}
4 g_{n+1/2}g_{n-1/2}\,A_{n}^2=g_{n-1}g_{n+1}B_{n}B_{n-1}+e_1(\lambda_n+\lambda_{n-1})-(e_2+q).
\end{align}
It is directly verified that \eqref{Equality-8} indeed satisfies the recurrence relation \eqref{Equality-6}. In addition, a straightforward calculation confirms that with \eqref{Actions-2}, \eqref{K1-Eigenvalues}, \eqref{Equality-8} and \eqref{Bn-Gn}, the defining relations \eqref{QR(3)} of the  Racah-Wilson algebra  are satisfied. The formula \eqref{Equality-8} for the matrix elements $A_n^2$ can be cast in the form
\begin{align}
\label{An-Poly}
A_{n}^2=\frac{\mathcal{P}(g_n^2)}{64\,g_n^2\,g_{n-1/2}\,g_{n+1/2}},
\end{align}
where $\mathcal{P}(z)$ is the fourth degree polynomial
\begin{align}
\begin{aligned}
\mathcal{P}(z)&=z^{4}-(4d+2)z^{3}+(4d^2+4d+1+8e_2-16e_1)z^{2}\\
&-4(d^2+2e_2+4de_2+4q)z+16e_2^2.
\end{aligned}
\end{align}
The polynomial $\mathcal{P}(g_n^2)$ is referred to as the characteristic polynomial of the algebra, as it determines the representation. Denoting by $\xi_j^2$, $j=1,\ldots,4$, the roots of this polynomial, one may write
\begin{align}
\label{An-Final}
A_{n}^2=\frac{\prod_{j=1}^{4}(g_n^2-\xi_j^2)}{64\,g_n^2\,g_{n-1/2}\,g_{n+1/2}}.
\end{align}
Using Vieta's formula for the roots of quartic polynomials \cite{Algebra}, one finds that the structure parameters $d$, $e_1$, $e_2$ and the Casimir value $q$ are related to the roots $\xi_k^{2}$ by
\begin{gather}
\begin{aligned}
\label{Parameters}
e_1&=\frac{1}{64}\big\{S_1^2-4S_{2}+ 8S_4^{1/2}\big\},\quad &&e_2=\frac{S_4^{1/2}}{4},\\
d&=\frac{1}{4}\big\{S_1-2\big\},&&q=\frac{1}{64}\big\{4S_1(1+ S_4^{1/2})+4S_3-S_1^2-4\big\},
\end{aligned}
\end{gather}
where $S_1,\cdots S_4$ are the elementary symmetric polynomials
\begin{align}
S_N=\sum_{1\leqslant i_1<i_2< \cdots < i_{N}\leqslant N}\xi_{i_1}^2\cdots \xi_{i_N}^2.
\end{align}
For real structure parameters, the roots $\xi_{k}^2$ are, in general, complex. Furthermore, one sees from \eqref{Parameters} that the structure parameters remain invariant under any transposition of the roots $\xi_k^2$ or change of sign of an even number of $\xi_k$.  The explicit formula for $A_{n}^2$ therefore reads
\begin{align}
\label{An-Roots}
A_{n}^2=\frac{1}{4}\frac{\prod_{j=1}^{4}(n-\sigma-\xi_{j})(n-\sigma+\xi_{j})}{(2n-2\sigma)^2(2n-2\sigma+1)(2n-2\sigma-1)}.
\end{align}
To obtain a finite-dimensional irreducible representation of \eqref{QR(3)}, one must have
\begin{align}
N_1\leqslant n\leqslant N_2,
\end{align}
where $N_1$, $N_2$ are integers. It is always possible to choose $N_1=0$ using the arbitrariness in the parameter $\sigma$ appearing in \eqref{K1-Eigenvalues}. Thus for $(N+1)$-dimensional irreducible representations to occur, the following conditions must hold:
\begin{align}
\label{Rel-5}
A_0=0,\quad A_{N+1}=0.
\end{align}
As is seen from \eqref{An-Final}, the truncation conditions \eqref{Rel-5} can be fulfilled if one has
\begin{align}
\label{Rel-3}
g_0=\pm \xi_i,\quad g_{N+1}=\pm \xi_{j},
\end{align}
where in the generic case $\xi_i\neq \xi_j$. The condition \eqref{Rel-3} implies that for finite-dimensional representations, at least two of the roots are real. The positivity condition $A_{n}^2>0$ induces conditions  on the possibles values for the roots $\xi_k$. As an example, consider the case for which the truncation conditions \eqref{Rel-5} are satisfied through
\begin{align}
\xi_1=\sigma,\quad \xi_4=(\sigma-N-1).
\end{align}
Then from \eqref{An-Roots}, it follows that $A_{n}^2>0$ for $n=1,\ldots,N$ if
\begin{align}
(\sigma<1/2\text{ or }\sigma>N+1/2)\text{ and }\xi_2^2<(\sigma-1)^2\text{ and }\xi_3^2>(\sigma-N)^2.
\end{align}
Similar conditions can be found by using the freedom in permuting and changing the signs of the $\xi_k$. Note that the with the condition on $\sigma$, the spectrum \eqref{K1-Eigenvalues} of $K_1$ is non-degenerate.

\subsection{Racah polynomials}
The relation between the finite-dimensional irreducible representations of the Racah-Wilson algebra and the classical Racah polynomials is as follows. Let $\phi_{s}$ be an eigenvector of $K_2$:
\begin{align}
\label{Dompe-1}
K_2\phi_{s}=\mu_{s}\phi_{s}.
\end{align}
The duality property of the algebra \eqref{QR(3)} can be used to derive the expression for the spectrum $\mu_{s}$ of $K_2$. Indeed, it is easily seen that the Racah-Wilson algebra is left invariant if one takes $\widetilde{K}_1=K_2$, $\widetilde{K}_2=K_1$, $\widetilde{K}_3=-K_3$ and performs the replacement $e_1\leftrightarrow e_2$. From this duality relation and \eqref{K1-Eigenvalues}, it follows that $\mu_s$ has the expression
\begin{align}
\mu_s=-(s-\nu)(s-\nu+1)/2,
\end{align}
where $\nu$ is an arbitrary parameter. In this basis, it follows from the duality that $K_1$ is three-diagonal with matrix elements given by the formulas \eqref{Bn-Gn} and \eqref{An-Poly} with $n\rightarrow s$ and $e_1\leftrightarrow e_2$.  

It is appropriate to indicate here the connection with Leonard pairs, which are defined as follows \cite{Terwilliger-2001}. Two linear transformations $(A_1, A_2)$ form a \emph{Leonard pair} if there exists a basis in which $A_1$ is diagonal and $A_2$ is tridiagonal and another basis in which $A_1$ is tridiagonal and $A_2$ is diagonal. The preceeding discussion makes clear that $K_1$, $K_2$ are hence realizing a Leonard pair.

 Since the two sets of basis vectors $\{\phi_s\}$ and $\{\psi_{n}\}$ for $n,s=0,\ldots,N$ span the same vector space, the two bases are related to one another by a linear transformation
\begin{align}
\label{Decomposition}
\phi_{s}=\sum_{n=0}^{N}W_{n}(s)\,\psi_{n}.
\end{align}
Upon acting with $K_2$ on each side of \eqref{Decomposition} and writing $W_{n}(s)=W_{0}(s)P_{n}(\mu_s)$ with $P_{0}(\mu_s)=1$, one finds that $P_{n}(\mu_s)$ satisfies
\begin{align}
\mu_sP_{n}(\mu_s)=A_{n+1}P_{n+1}(\mu_s)+B_{n}\psi_{n}(\mu_s)+A_{n}P_{n-1}(\mu_s).
\end{align}
Hence $P_{n}(\mu_s)$ are polynomials of degree $n$ in the variable $\mu_s$. The above recurrence relation can be put in monic form by taking $P_{n}(\mu_s)=(A_1\cdots A_{n})^{-1}\widehat{P}_{n}(x)$. One then has
\begin{align}
\label{Monic}
x\widehat{P}_{n}(x)=\widehat{P}_{n}(x)+B_{n}\widehat{P}_{n}(x)+A_{n}^2\widehat{P}_{n-1}(x).
\end{align}
It follows from \eqref{Monic} that the polynomials $\widehat{P}_{0}(x),\,\widehat{P}_1(x)\ldots,\,\widehat{P}_{N}(x)$ form a finite system of orthogonal polynomials provided that $A_{n}^2>0$ for $n=1,\ldots,N$ \cite{Chihara-1978}. It can be seen that these polynomials correspond to the classical Racah polynomials. To obtain the identification, one first introduces the monic polynomials $\widehat{H}_{n}(\widetilde{x})$ in the variable $\widetilde{x}=-2(x+\tau)$ by taking $\widehat{P}_{n}(x)=(-2)^{-n}\widehat{H}_{n}(\widetilde{x})$. Upon using the formulas \eqref{Parameters} for the structure parameters in terms of the roots $\xi_k^2$ and the explicit expressions \eqref{K1-Eigenvalues}, \eqref{Bn-Gn}  and \eqref{An-Roots} for $\lambda_{n}$, $B_{n}$ and $A_{n}^2$, one finds that the polynomials $\widehat{H}_{n}(\widetilde{x})$ obey the recurrence relation
\begin{align}
\label{Recu-2}
\widetilde{x}\widehat{H}_{n}(\widetilde{x})=\widehat{H}_{n+1}(\widetilde{x})+\widetilde{B}_{n}\widehat{H}_{n}(\widetilde{x})+\widetilde{A}_{n}^2\widehat{H}_{n-1}(\widetilde{x}),
\end{align}
where 
\begin{align}
\begin{aligned}
\label{Recu-Coef}
\widetilde{B}_{n}&=\frac{1}{2}(\sigma-n)(n-\sigma+1)+\frac{\xi_1\xi_2\xi_3\xi_4}{2(\sigma-n)(n-\sigma+1)}+\frac{1}{4}\sum_{j=1}^{4}\xi_j^2-(2\tau+1/2),\\
\widetilde{A}_{n}^2&=\frac{\prod_{j=1}^{4}((\sigma-n)^2-\xi_{j}^2)}{(2n-2\sigma)^2(2n-2\sigma+1)(2n-2\sigma-1)}.
\end{aligned}
\end{align}
The recurrence relation \eqref{Recu-2} and recurrence coefficients \eqref{Recu-Coef} can now be compared with those of the monic Racah polynomials $\widehat{R}_{n}(\lambda(x);\alpha,\beta,\gamma,\delta)$. These polynomials are defined by \cite{Koekoek-2010}
\begin{align}
\widehat{R}_{n}&(\lambda(x);\alpha,\beta,\gamma,\delta)=
\nonumber
\\
\label{Def-Racah}
&\frac{(\alpha+1)_{n}(\beta+\delta+1)_{n}(\gamma+1)_{n}}{(n+\alpha+\beta+1)_{n}}\; \pFq{4}{3}{-n,n+\alpha+\beta+1,-x,x+\gamma+\delta+1}{\alpha+1,\beta+\delta+1,\gamma+1}{1}
\end{align}
where $(a)_{n}=a(a+1)\cdots (a+n-1)$ denotes the Pochhammer symbol and where ${}_pF_{q}$ stands for the generalized hypergeometric function \cite{Gasper-2004}. The monic Racah polynomials obey the three-term recurrence relation \cite{Koekoek-2010}
\begin{align}
\label{Recu-Racah}
x\,\widehat{R}_{n}(\lambda(x))=\widehat{R}_{n+1}(\lambda(x))-(C_{n}+D_{n})\widehat{R}_{n}(\lambda(x))+C_{n-1}D_{n}\widehat{R}_{n}(\lambda(x)),
\end{align}
where the recurrence coefficients are given by
\begin{align}
\begin{aligned}
C_{n}&=\frac{(n+\alpha+1)(n+\alpha+\beta+1)(n+\gamma+1)(n+\beta+\delta+1)}{(2n+\alpha+\beta+1)(2n+\alpha+\beta+2)},\\
D_{n}&=\frac{n(n+\beta)(n+\alpha-\delta)(n+\alpha+\beta-\gamma)}{(2n+\alpha+\beta)(2n+\alpha+\beta+2)},
\end{aligned}
\end{align}
and where it assumed that one of the conditions $\alpha+1=-N$, $\beta+\delta+1=-N$ or $\gamma+1=-N$ holds. Now suppose that the truncation conditions $A_0=0$, $A_{N+1}=0$ are satisfied in \eqref{Recu-Coef} through
\begin{align}
g_0=\sigma=\xi_1,\quad g_{N+1}=(\sigma-N-1)=\xi_4.
\end{align}
Then one can adopt the following parametrization for the roots:
\begin{align}
\label{dompe-2}
\xi_1=-\frac{\alpha+\beta}{2},\quad \xi_2=\frac{\beta-\alpha}{2}+\delta,\quad \xi_3=\frac{\beta-\alpha}{2},\quad \xi_4=\gamma-\frac{\alpha+\beta}{2},
\end{align}
for which one has $\gamma+1=-N$. Then with $\tau=(2+\gamma+\delta)(\gamma+\delta)/8$, the recurrence relation \eqref{Recu-2} is directly seen to coincide with \eqref{Recu-Racah}. It is clear from the above considerations that the parameters $\xi_i$, $i=1,\ldots,4$, are much more convenient for the analysis in terms of orthogonal polynomials.
\subsection{Realization of the Racah algebra}
The bispectrality of the Racah polynomials can be used to obtain a realization of the reduced Racah-Wilson algebra \eqref{QR(3)} in terms of difference operators. The bispectral property of the Racah polynomials \eqref{Def-Racah} is as follows. On the one hand, the polynomials $R_{n}(\lambda(x))$ satisfy the three term recurrence relation
\begin{align}
\label{Recu-3}
\lambda(x) R_{n}(\lambda(x))=C_{n}R_{n+1}(\lambda(x))-(C_{n}+D_{n})R_{n}(\lambda(x))+D_{n}R_{n-1}(\lambda(x)),
\end{align}
where $C_{n}$ and $D_{n}$ are given by \eqref{Recu-Racah} and where $\lambda(x)=x(x+\gamma+\delta+1)$. On the other hand, the polynomials also satisfy the eigenvalue equation
\begin{align}
\label{Diff-1}
\big[B(x)T^{+}-(B(x)+E(x))+E(x) T^{-}\big]R_{n}(\lambda(x))=\mu_n R_{n}(\lambda(x)),
\end{align}
with $\mu_n=n(n+\alpha+\beta+1)$ and where $T^{+}f(x)=f(x+1)$, $T^{-}f(x)=f(x-1)$ are the usual shift operators. The coefficients appearing in \eqref{Diff-1} are given by
\begin{align}
\begin{aligned}
B(x)&=\frac{(x+\alpha+1)(x+\beta+\delta+1)(x+\gamma+1)(x+\gamma+\delta+1)}{(2x+\gamma+\delta+1)(2x+\gamma+\delta+2)},\\
E(x)&=\frac{x(x-\alpha+\gamma+\delta)(x-\beta+\gamma)(x+\delta)}{(2x+\gamma+\delta)(2x+\gamma+\delta+1)}.
\end{aligned}
\end{align}
The recurrence operator  \eqref{Recu-3} and the difference operator \eqref{Diff-1} can be used to realize the Racah-Wilson algebra. To this end, the recurrence operator is denoted $K_1$ and taken to be diagonal, i.e. as in the LHS of \eqref{Recu-3}. The difference operator is denoted $K_2$ and taken to be three-diagonal, i.e. the LHS of \eqref{Diff-1}. Upon introducing $K_3=[K_1,K_2]$, one thus has
\begin{align}
\begin{aligned}
K_1&=x(x+\gamma+\delta+1),\\
K_2&=B(x)T^{+}+E(x)T^{-}-(B(x)+E(x)),\\
K_3&=(2x+\gamma+\delta)E(x)T^{-}-(2x+\gamma+\delta+2)B(x) T^{+}.
\end{aligned}
\end{align}
A direct computation shows that the operators $K_1$, $K_2$ and $K_3$ realize the Racah-Wilson algebra \eqref{QR} with structure parameters
\begin{gather*}
a_1=-2,\quad a_2=-2\\
 c_1=-(\alpha+\beta)(2+\alpha+\beta),\quad c_2=-(\gamma+\delta)(2+\gamma+\delta),\\
e_1=-(\alpha+1)(\alpha+\beta)(\beta+\delta+1)(\gamma+1),\quad e_2=-(\alpha+1)(\beta+\delta+1)(\gamma+1)(\gamma+\delta),\\
d=\beta(\delta-\gamma-2)-\alpha(2\beta+\gamma+\delta+2)-2(\gamma+1)(\delta+1).
\end{gather*}
The canonical form \eqref{QR(3)} can be obtained if one takes
\begin{align}
K_1\rightarrow a_2^{-1}K_1+c_2/2a_2^2,\quad K_2\rightarrow a_1^{-1}K_2+c_1/2a_1^2,\quad K_3\rightarrow a_1^{-1}a_2^{-1}K_3.
\end{align}
The remaining non-zero structure constants become
\begin{align}
\begin{aligned}\label{above}
e_1&\rightarrow\frac{1}{4}
\left(\frac{\alpha-\beta}{2}\right)
\left(\frac{\alpha+\beta}{2}\right)
\left(\frac{\alpha+\beta}{2}-\gamma\right)
\left(\frac{\alpha-\beta}{2}-\delta\right),
\\
e_2&\rightarrow\frac{1}{4}
\left(\frac{\gamma-\delta}{2}\right)\left(\frac{\gamma+\delta}{2}\right)
\left(\frac{\gamma+\delta}{2}-\alpha\right)
\left(\frac{\gamma-\delta}{2}-\beta\right),
\\
d&\rightarrow \frac{1}{4}
\left\{
\left(\frac{\gamma-\delta}{2}\right)^2
+\left(\frac{\gamma+\delta}{2}\right)^2
+\left(\frac{\gamma+\delta}{2}-\alpha\right)^2
+\left(\frac{\gamma-\delta}{2}-\beta\right)^2-2
\right\}\\
\quad &= \frac{1}{4}
\left\{
\left(\frac{\alpha-\beta}{2}\right)^2
+\left(\frac{\alpha+\beta}{2}\right)^2
+\left(\frac{\alpha+\beta}{2}-\gamma\right)^2
+\left(\frac{\alpha-\beta}{2}-\delta\right)^2-2
\right\}.
\end{aligned}
\end{align}
It is interesting to note that the duality property of the Racah polynomials is encoded in the Racah-Wilson algebra \eqref{QR(3)} and can be derived directly from the structure constants \eqref{above}. Indeed, the algebra is invariant under the exchange $K_1\leftrightarrow K_2$, $K_3\rightarrow -K_3$ and $e_1\leftrightarrow e_2$. This means that if one instead takes $K_1$ as the diagonal operator in the RHS of \eqref{Diff-1} and  $K_2$ as the tridiagonal operator in the RHS of \eqref{Recu-3}, then one finds the same algebra with $e_1\leftrightarrow e_2$. From \eqref{above}, this is equivalent to the well-known duality property of the Racah polynomials ($n\leftrightarrow x$, $\alpha\leftrightarrow \gamma$, $\beta\leftrightarrow \delta$):
\begin{align}
R_{n}(\lambda(x);\alpha,\beta,\gamma,\delta)=R_{x}(\lambda(n);\gamma,\delta,\alpha,\beta).
\end{align}

\section{The Racah problem for $\mathfrak{su}(1,1)$ and the Racah-Wilson algebra}
In this section, the Racah problem for $\mathfrak{su}(1,1)$ is revisited using the Racah-Wilson algebra. It shall be shown that the Racah-Wilson algebra is the fundamental algebraic structure behind this problem. Furthermore, it will be seen that our approach encompasses simultaneously the combination of three unitary irreducible $\mathfrak{su}(1,1)$ representations of any series.

\subsection{Racah problem essentials for $\mathfrak{su}(1,1)$}
The $\mathfrak{su}(1,1)$ algebra is generated by the elements $J_{0}$, $J_{\pm}$ with commutation relations
\begin{align}
\label{Relations}
[J_{0},J_{\pm}]=\pm J_{\pm},\quad [J_{+},J_{-}]=-2J_{0}.
\end{align}
The Casimir operator, which commutes with all $\mathfrak{su}(1,1)$ elements, is given by
\begin{align}
\mathcal{C}=J_0^2-J_{+}J_{-}-J_{0}.
\end{align}
Let $J_{0}^{(i)}$, $J_{\pm}^{(i)}$, $i=1,2,3$, denote three mutually commuting sets of generators satisfying the commutation relations \eqref{Relations}. The three sets can be combined to produce a fourth set of $\mathfrak{su}(1,1)$ generators as follows :
\begin{align}
\label{Algebra-4}
J_0^{(4)}=J_{0}^{(1)}+J_{0}^{(2)}+J_{0}^{(3)},\quad J_{\pm}^{(4)}=J_{\pm}^{(1)}+J_{\pm}^{(2)}+J_{\pm}^{(3)}.
\end{align}
The Casimir operator $\mathcal{C}^{(4)}=J_0^{(4)}-J_{+}^{(4)}J_{-}^{(4)}-J_{0}^{(4)}$ for the representation \eqref{Algebra-4} of $\mathfrak{su}(1,1)$ is easily seen to have the following expression:
\begin{align}
\label{Full-Casimir}
\mathcal{C}^{(4)}=\mathcal{C}^{(12)}+\mathcal{C}^{(23)}+\mathcal{C}^{(31)}-\mathcal{C}^{(1)}-\mathcal{C}^{(2)}-\mathcal{C}^{(3)},
\end{align}
where $\mathcal{C}^{(i)}=J_0^{(i)}-J_{+}^{(i)}J_{-}^{(i)}-J_{0}^{(i)}$ are the individual Casimir operators and $\mathcal{C}^{(ij)}$ are the intermediate Casimir operators
\begin{align}
\label{Intermediate}
\mathcal{C}^{(ij)}=2J_0^{(i)}J_{0}^{(j)}-(J_{+}^{(i)}J_{-}^{(j)}+J_{+}^{(j)}J_{-}^{(j)})+\mathcal{C}^{(i)}+\mathcal{C}^{(j)}.
\end{align}
The full Casimir operator $\mathcal{C}^{(4)}$ commutes with all the intermediate Casimir operators $\mathcal{C}^{(ij)}$ and with all the individual Casimir operators $\mathcal{C}^{(i)}$. The intermediate Casimir operators $\mathcal{C}^{(ij)}$ do not commute with one another but commute with each of the individual Casimir operators $\mathcal{C}^{(i)}$ and with $\mathcal{C}^{(4)}$.

The Racah problem can be posited as follows. Let $V^{(\lambda_i)}$, $i=1,2,3$, denote a generic unitary irreducible representation space on which the Casimir operator $\mathcal{C}^{(i)}$ has the eigenvalue $\lambda_i$. The irreducible unitary representations of $\mathfrak{su}(1,1)$ are known and classified (see for example \cite{Klimyk-1991}). Note that $V^{(\lambda_i)}$ may depend on additional parameters other than $\lambda_i$ and that $V^{(\lambda_i)}$ need not be of the same type as $V^{(\lambda_j)}$. A representation space $V$ for the algebra \eqref{Algebra-4} is obtained by taking $V=V^{(\lambda_1)}\otimes V^{(\lambda_2)}\otimes
V^{(\lambda_3)}$, where it is understood that each set of $\mathfrak{su}(1,1)$ generators $J_0^{(i)}$, $J_{\pm}^{(i)}$ acts on the corresponding representation space $V^{(\lambda_i)}$. In general, the representation space $V$ is not irreducible and can be decomposed into irreducible components in two equivalent ways.
\begin{itemize}
\item In the first scheme, one decomposes $V^{(\lambda_1)}\otimes V^{(\lambda_2)}$ in irreducible components $V^{(\lambda_{12})}$ and then further decomposes $V^{(\lambda_{12})}\otimes V^{(\lambda_3)}$ for each occurring values of $\lambda_{12}$. On the spaces $V^{(\lambda_{12})}$, the intermediate Casimir operator $\mathcal{C}^{(12)}$ acts as $\lambda_{12}\cdot \unit$.

\item In the second scheme, one first decomposes $V^{(\lambda_2)}\otimes V^{(\lambda_3)}$ in irreducible components $V^{(\lambda_{23})}$ and then further decompose $V^{(\lambda_{1})}\otimes V^{(\lambda_{23})}$ for each occurring values of $\lambda_{23}$. On the space $V^{(\lambda_{23})}$, the intermediate Casimir operator $\mathcal{C}^{(23)}$ acts as $\lambda_{23}\cdot \unit$.
\end{itemize}
One can define two natural orthonormal bases for the representation space $V$ which correspond to the two different decomposition schemes. For the first scheme, the natural orthonormal basis vectors that span $V$ are denoted $\ket{\lambda_{12};\vec{\lambda}}$ and are defined by
\begin{align}
\mathcal{C}^{(12)}\ket{\lambda_{12};\vec{\lambda}}=\lambda_{12}\ket{\lambda_{12};\vec{\lambda}},\quad \mathcal{C}^{(i)}\ket{\lambda_{12};\vec{\lambda}}=\lambda_{i}\ket{\lambda_{12};\vec{\lambda}},
\end{align}
where $\vec{\lambda}=(\lambda_1,\lambda_2,\lambda_3,\lambda_4)$. For the second scheme, the natural orthonormal basis vectors are denoted $\ket{\lambda_{23};\vec{\lambda}}$ and are defined by
\begin{align}
\mathcal{C}^{(23)}\ket{\lambda_{23};\vec{\lambda}}=\lambda_{23}\ket{\lambda_{23};\vec{\lambda}},\quad \mathcal{C}^{(i)}\ket{\lambda_{23};\vec{\lambda}}=\lambda_{i}\ket{\lambda_{23};\vec{\lambda}}.
\end{align}
The three parameters $\lambda_1$, $\lambda_2$ and $\lambda_3$ are given while $\lambda_{12}$, $\lambda_{23}$ and $\lambda_{4}$ vary so that the basis vectors span $V$. The possible values for these parameters depend on the representations $V^{(\lambda_i)}$ that are involved in the tensor product. For a given value of $\lambda_4$, the orthonormal vectors $\ket{\lambda_{12};\vec{\lambda}}$, $\ket{\lambda_{23};\vec{\lambda}}$ with admissible values of $\lambda_{12}$, $\lambda_{23}$  span the same space and are thus related by a unitary transformation. One can thus write
\begin{align}
\label{Decompo-Racah}
\ket{\lambda_{12};\vec{\lambda}}=\sum_{\lambda_{23}}\braket{\lambda_{23};\vec{\lambda}}{\lambda_{12};\vec{\lambda}}\;\ket{\lambda_{23};\vec{\lambda}}=\sum_{\lambda_{23}}R^{\vec{\lambda}}_{\lambda_{12},\lambda_{23}}\ket{\lambda_{23};\vec{\lambda}},
\end{align}
where the range of the sum depends on the possible values for $\lambda_{23}$. Note that these values (or those of $\lambda_{12}$) may vary continuously hence the sum in \eqref{Decompo-Racah} can also be an integral. The expansion coefficients $\braket{\lambda_{23};\vec{\lambda}}{\lambda_{12};\vec{\lambda}}=R_{\lambda_{12},\lambda_{23}}^{\vec{\lambda}}$ between the two bases with vectors $\ket{\lambda_{12};\vec{\lambda}}$ and $\ket{\lambda_{23};\vec{\lambda}}$ are known as Racah coefficients. These coefficients are usually taken to be real. Since the two bases are orthonormal, the Racah coefficients satisfy the orthogonality relations
\begin{align}
\sum_{\lambda_{23}}R^{(\lambda)}_{\lambda_{12},\lambda_{23}}R^{(\lambda)}_{\lambda_{12}',\lambda_{23}}=\delta_{\lambda_{12}\lambda_{12}'},\quad \sum_{\lambda_{12}}R^{(\lambda)}_{\lambda_{12},\lambda_{23}}R^{(\lambda)}_{\lambda_{12},\lambda_{23}'}=\delta_{\lambda_{23}\lambda_{23}'},
\end{align}
where each of the summation may become an integral depending on the possible values of $\lambda_{12}$, $\lambda_{23}$. The evaluation of the coefficients in \eqref{Decompo-Racah} is referred to as the ``Racah problem''.
\subsection{The Racah problem and the Racah-Wilson algebra}
It will now be shown that the Racah-Wilson algebra is the fundamental structure behind the Racah problem. The idea behind the method is the following. Since the vectors $\ket{\lambda_{12};\vec{\lambda}}$ and $\ket{\lambda_{23};\vec{\lambda}}$ are the eigenvectors of the two non-commuting intermediate Casimir operators $\mathcal{C}^{(12)}$ and $\mathcal{C}^{(23)}$, respectively, the information on the structure of the coefficients $\braket{\lambda_{23};\vec{\lambda}}{\lambda_{12};\vec{\lambda}}$ can be obtained by studying their commutation relations. The Casimir operators $\mathcal{C}^{(i)}$, $i=1,\ldots,4$, all act as multiples of the identity on both sets of vectors $\ket{\lambda_{12},\vec{\lambda}}$, $\ket{\lambda_{23},\vec{\lambda}}$. Consequently, they can be treated as constants, i.e.:
\begin{align}
\label{Cas-Const}
\mathcal{C}^{(i)}=\lambda_i,\quad i=1,\ldots,4.
\end{align}
Let $\kappa_1$ and $\kappa_2$ be defined as
\begin{align}
\kappa_1=-\frac{1}{2}\,\mathcal{C}^{(12)},\quad \kappa_2=-\frac{1}{2}\,\mathcal{C}^{(23)}.
\end{align}
Using the identification \eqref{Cas-Const} and the definition \eqref{Intermediate}, a direct computation shows that $\kappa_1$, $\kappa_2$ and their commutator 
\begin{align*}
[\kappa_1,\kappa_2]=\kappa_3=\frac{J_0^{(1)}}{2}\left(J_{-}^{(2)}J_{+}^{(3)}-J_{+}^{(2)}J_{-}^{(3)}\right)+\text{cyclic permutations},
\end{align*}
satisfy the commutation relations of the Racah-Wilson algebra \eqref{QR(3)}
\begin{align}
\begin{aligned}\label{Result}
[\kappa_1,\kappa_2]&=\kappa_3,\\
[\kappa_2,\kappa_3]&=\kappa_2^2+\{\kappa_1,\kappa_2\}+d\,\kappa_2+e_1,\\
[\kappa_3,\kappa_1]&=\kappa_1^2+\{\kappa_1,\kappa_2\}+d\,\kappa_1+e_2,
\end{aligned}
\end{align}
with structure parameters
\begin{subequations}
\label{Structure-Parameters}
\begin{gather}
d=\frac{1}{2}(\lambda_1+\lambda_2+\lambda_3+\lambda_4),\\
e_1=\frac{1}{4}(\lambda_1-\lambda_4)(\lambda_2-\lambda_3),\quad e_2=\frac{1}{4}(\lambda_1-\lambda_2)(\lambda_4-\lambda_3).
\end{gather}
\end{subequations}
It is thus seen that the Racah-Wilson algebra is the fundamental algebraic structure behind the Racah problem for $\mathfrak{su}(1,1)$. One recalls that this result is valid for any choice of representations corresponding to $V^{(\lambda_i)}$ provided that the Casimir operator acts with eigenvalue $\lambda_i$. This indicates that it is possible to obtain all types of Racah coefficients corresponding to the combination of the various $\mathfrak{su}(1,1)$ unitary representations through the analysis of the representations of the Racah-Wilson algebra (see for example \cite{Groenevelt-2005,Koelink-2003} for possible applications of this scheme). We also note that a similar result holds for the combination of two $\mathfrak{su}(1,1)$ irreducible representations (Clebsch-Gordan problem), where the algebra \eqref{QR} appears with $a_2\cdot a_1=0$. In this case the relevant operators are the intermediate Casimir operator $\kappa_1=\mathcal{C}^{(12)}$ and the operator $\kappa_2=J_0^{(1)}-J_{0}^{(2)}$.
\subsection{Racah problem for the positive-discrete series}
The above results will now be specialized to the positive discrete series of unitary representations of $\mathfrak{su}(1,1)$; these representations will occur in the correspondence between the $\mathfrak{su}(1,1)$ Racah problem and the analysis of the generic 3-parameter superintegrable system on the two sphere. The positive discrete series of unitary irreducible representations of $\mathfrak{su}(1,1)$ are infinite-dimensional and labeled by a positive real number $\nu$. They can be defined by the following actions on a canonical basis $\ket{\nu,n}$, $n\in \mathbb{N}$:
\begin{align}
\begin{aligned}\label{Positive-Discrete}
J_0\ket{\nu,n}&=(n+\nu)\,\ket{\nu,n},\\
J_{+}\ket{\nu,n}&=\sqrt{(n+1)(n+2\nu)}\,\ket{\nu,n+1},\\
J_{-}\ket{\nu,n}&=\sqrt{n(n+2\nu-1)}\,\ket{\nu,n-1}.
\end{aligned}
\end{align}
The action of the Casimir operator $\mathcal{C}$ is given by
\begin{align}
\label{para}
\mathcal{C}\ket{\nu,n}=\nu(\nu-1)\ket{\nu,n}.
\end{align}
Let us now consider the Racah problem for the combination of three representations of the discrete series, each labeled by a positive number $\nu_i$, $i=1,2,3$. In this case, the structure constants in the algebra \eqref{Result} have the following expressions:
\begin{align}
\label{Dompe-4}
\lambda_i=\nu_i(\nu_i-1),\quad i=1,\ldots,4.
\end{align}
With the eigenvalues of the Casimir operators parametrized as in \eqref{Dompe-4}, we will replace the notation $\ket{\lambda_{ij};\vec{\lambda}}$ for instance by $\ket{\nu_{ij};\vec{\nu}}$.
There remains to evaluate the admissible values of $\nu_{12}$, $\nu_{23}$ and $\nu_4$. We begin with $\nu_{12}$. In view of the addition rule $J_0^{(12)}=J_{0}^{(1)}+J_{0}^{(2)}$ and the actions \eqref{Positive-Discrete}, it is not hard to see that the possible values of $\nu_{12}$ are of the form
\begin{align}
\label{Intermediate-Values-1}
\nu_{12}=\nu_1+\nu_{2}+n_{12},\quad n_{12}\in \mathbb{N}.
\end{align}
Similarly, one has
\begin{align}
\label{Intermediate-Values-2}
\nu_{23}=\nu_2+\nu_3+n_{23},\quad n_{23}\in \mathbb{N}. 
\end{align}
Again, the addition rule for $J_0^{(4)}=J_{0}^{(12)}+J_{0}^{(3)}=J_0^{(1)}+J_{0}^{(23)}$ gives for $\nu_4$
\begin{align}
\label{Full-Casimir-Values}
\nu_4=\nu_1+\nu_2+\nu_3+N=\nu_{12}+\nu_{3}+p_1=\nu_1+\nu_{23}+p_2,
\end{align}
where $N,p_1,p_2\in \mathbb{N}$. For a given of $\nu_4$, the dimension of the space spanned by the basis vectors $\ket{\nu_{12};\vec{\nu}}$ can be evaluated from \eqref{Full-Casimir-Values} and \eqref{Intermediate-Values-1} in the following way. If $N=0$, it is obvious that there is only one possible value for $\nu_{12}$. If $N=1$, then $\nu_4=\nu_1+\nu_2+\nu_3+1$ and hence $\nu_{12}$ can take two values corresponding to $n_{12}=0,1$. By a direct inductive argument, the dimension of the space spanned by the basis vectors  $\ket{\nu_{12};\vec{\nu}}$ is $N+1$.

Let us return to the results of Section 2. Upon comparing the structure constants in \eqref{Result} with \eqref{Parameters} and using \eqref{Dompe-4}, it is seen that the $\xi_1,\ldots \xi_4$ of the characteristic polynomial of the Racah-Wilson algebra can be taken to be
\begin{align}
\label{Parameters-2}
\xi_1=(1-\nu_1-\nu_2),\quad \xi_2=(\nu_1-\nu_2),\quad \xi_3=(\nu_4+\nu_3-1),\quad \xi_4=(\nu_3-\nu_4).
\end{align}
The free parameter $\sigma$ in the spectrum of $\kappa_1$ given by \eqref{K1-Eigenvalues} is evaluated to 
\begin{align}
\sigma=1-\nu_1-\nu_2,
\end{align}
since the minimal value of $\nu_{12}$ is $\nu_1+\nu_2$. From \eqref{Parameters-2} and \eqref{Full-Casimir-Values}, it is seen that the truncation conditions
\begin{align}
\xi_1=\sigma,\quad \xi_4=(\sigma-N-1),
\end{align}
are satisfied. Thus it follows that the Racah coefficients for the combination of three $\mathfrak{su}(1,1)$ representations of the positive discrete series are expressed in terms of the Racah polynomials with the parameter identification obtained by combining \eqref{Parameters} and \eqref{Parameters-2}.
\section{The 3-parameter superintegrable system on the 2-sphere and the $\mathfrak{su}(1,1)$ Racah problem}
The stage has now been set to establish the equivalence between the Racah problem for $\mathfrak{su}(1,1)$ and the analysis of the generic 3-parameter superintegrable system of the two-sphere. To this end, consider the following differential realizations of $\mathfrak{su}(1,1)$
\begin{align}
\begin{aligned}
\label{Realization}
J_0^{(i)}&=\frac{1}{4}\left(-\pd_{x_i}^2+x_i^2+\frac{(k_i^2-1/4)}{x_i^2}\right),\\
J_{\pm}^{(i)}&=\frac{1}{4}\left(\pd_{x_i}^2\mp 2x_i\pd_{x_i}+(x_i^2\mp 1)-\frac{(k_i^2-1/4)}{x_i^2}\right),
\end{aligned}
\end{align}
where $i=1,2,3$. In these realizations, the Casimir operators $\mathcal{C}^{(i)}$ have actions:
\begin{align*}
\mathcal{C}^{(i)}f(x_i)=\nu_i(\nu_i-1)f(x_i),
\end{align*}
where $\nu_i=(k_{i}+1)/2$. It is thus seen that $\nu_i>0$ if $k_i>-1$. The operator $J_0^{(i)}$ is the Hamiltonian of the singular oscillator and it has a positive and discrete spectrum. Hence the representation \eqref{Realization} realize the positive discrete series of $\mathfrak{su}(1,1)$. Upon using \eqref{Realization}, it is observed that the full Casimir operator $\mathcal{C}^{(4)}$ and the intermediate Casimir operators $\mathcal{C}^{(ij)}$ have the expressions
\begin{align}
\begin{aligned}
\mathcal{C}^{(4)}&=\frac{1}{4}\Big\{J_1^2+J_2^2+J_3^2+(x_1^2+x_2^2+x_3^2)\left(\frac{a_1}{x_1^2}+\frac{a_2}{x_2^2}+\frac{a_3}{x_3^2}\right)-\frac{3}{4}\Big\},\\
\mathcal{C}^{(ij)}&=\frac{1}{4}\Big\{J_k^2+\frac{a_ix_j^2}{x_i^2}+\frac{a_jx_i^2}{x_j^2}+a_i+a_j-1\Big\},
\end{aligned}
\end{align}
where the indices $i$, $j$, $k$ are such that $\epsilon_{ijk}=1$, $a_i=k_i^2-1/4$ and where the operators $J_i$ are the angular momentum generators \eqref{Def-J}. Upon returning to the defining formulas \eqref{Def-Symmetries} for the symmetries of the generic 3-parameter system on the two-sphere, it is directly seen that
\begin{align}
L_{k}=4\,\mathcal{C}^{(ij)}-a_i-a_j+1,
\end{align}
where again the indices are such that $\epsilon_{ijk}=1$. It is also seen that
\begin{align}
\mathcal{H}=4\,\mathcal{C}^{(4)}+3/4,
\end{align}
if the condition $x_1^2+x_2^2+x_3^2=1$ is satisfied. This condition can be ensured in general. Indeed, it is verified that
\begin{align}
S=2J_0^{(4)}+J_{+}^{(4)}+J_{-}^{(4)}=x_1^2+x_2^2+x_3^2.
\end{align}
Since the operator $S$ commutes with $\mathcal{C}^{(4)}$ and all the intermediate Casimir operators $\mathcal{C}^{(ij)}$, it is central in the algebra \eqref{Result} and thus it can be considered as a constant. Consequently, one can take $S=1$ without loss of generality and this completes the identification of $\mathcal{H}$ with $\mathcal{C}^{(4)}$.

We have thus identified the full Casimir operator $\mathcal{C}^{(4)}$ of the combination of three $\mathfrak{su}(1,1)$ algebras with the Hamiltonian of the generic three-parameter superintegrable system on the two sphere and we have also identified the intermediate Casimir operators $\mathcal{C}^{(12)}$, $\mathcal{C}^{(23)}$ and $\mathcal{C}^{(31)}$ with the symmetries $L_3$, $L_1$, $L_2$, respectively, of this Hamiltonian. In view of the result \eqref{Result}, it follows that the symmetry algebra \eqref{Symmetry-Algebra-1}, \eqref{Symmetry-Algebra-2} of the generic 3-parameter system on the two sphere coincides with the Racah-Wilson algebra \eqref{QR(3)} with structure parameters \eqref{Structure-Parameters}. We also note that the conditions for the $\nu_i=(k_i+1)/2$ to be positive are the same conditions for the Hamiltonian $\mathcal{H}$ to have normalizable solutions. Moreover, the spectrum found for the full Casimir operator $\mathcal{C}^{(4)}$ yields for the energies (eigenvalues) of the Hamiltonian
\begin{align}
E_{N}=4\nu_4(\nu_4-1)+\frac{3}{4}=\big[2(N+1)+k_1+k_2+k_3\big]^2-\frac{1}{4},
\end{align}
where $N$ is a non-negative integer. This $(N+1)$-fold degenerate spectrum coincides, as should be, with the one obtained in \cite{Kalnins-1996} for the spectrum of the Hamiltonian of the generic 3-parameter system.
\section{Conclusion}
In this paper, it has been shown that the analysis of the most general second-order superintegrable system in two dimensions, i.e. the generic 3-parameter system on the 2-sphere, is equivalent to the Racah problem for the positive-discrete series of unitary representations of the Lie algebra $\mathfrak{su}(1,1)$. This correspondence establishes that the symmetry algebra \eqref{Symmetry-Algebra-1}, \eqref{Symmetry-Algebra-2} of the generic 3-parameter system is isomorphic to the reduced Racah-Wilson algebra \eqref{QR(3)}. Since the representations of the Racah-Wilson algebra are related to the Racah and Wilson polynomials, this provides an explanation for the connection between the Racah polynomials and the superintegrable  3-parameter system on the 2-sphere.

It has been shown that the Racah-Wilson algebra defining relations can also be realized \cite{Gao-2013} by taking $K_1$, $K_2$ and $K_3$ as quadratic expressions in the generators (in the equitable presentation) of one $\mathfrak{su}(2)$ algebra. It is relevant to understand how this construction pertains to the relation between the Racah-Wilson algebra and the composition of three $\mathfrak{su}(1,1)$ representations. In differential operator terms, this asks the question of how can one pass from a three- to a one-variable model. This will be explained in a forthcoming publication \cite{Zhedanov-2013}. Somewhat related would be the algebraic description of the tridiagonalization of ordinary and basic hypergeometric operators \cite{Koelink-2012}. Finally, it is of considerable interest to pursue the analysis of superintegrable models in 3 dimensions along the lines of the present paper. The relation between the generic model on the 3-sphere and Racah/Wilson polynomials in two variables has already been established \cite{Miller-2011}. It is quite clear that these polynomials should correspond to the $9j$ symbol of $\mathfrak{su}(1,1)$ and that the underlying algebra should describe the symmetries. This analysis should lift the veil on the study and standardization of polynomial algebras of ``rank two'' associated to bivariate orthogonal polynomials and superintegrable models in three dimensions. We plan to report on these questions in the near future.
\section*{Acknowledgements}
The authors wish to thank P. Terwilliger for stimulating discussions and S. Gao for providing a copy of the paper \cite{Gao-2013} in proofs. V.X.G. holds an Alexander-Graham-Bell fellowship from NSERC. The research of L.V. is supported in part by the Natural Sciences and Engineering Council of Canada (NSERC). 
\small

\begin{thebibliography}{10}

\bibitem{Chihara-1978}
T.~Chihara.
\newblock {\em {An {I}ntroduction to {O}rthogonal {P}olynomials}}.
\newblock Gordon and Breach, New-York, 1978.

\bibitem{Daska-2006}
C.~Daskaloyannis and K.~Ypsilantis.
\newblock {Unified treatment and classification of superintegrable systems with
  integrals quadratic in momenta on a two dimensonal manifold}.
\newblock {\em Journal of Mathematical Physics}, 47:042904, 2006.

\bibitem{Winter-1965}
I.~Fris, V.~Mandrosov, Y.~A. Smorodinsky, M.~Ulh{\`i}r, and P.~Winternitz.
\newblock {On higher symmetries in quantum mechanics}.
\newblock {\em Physics Letters}, 16:354--356, 1965.

\bibitem{Zhedanov-1991-2}
O.F. Gal'bert, Y.~Granovskii, and A.~Zhedanov.
\newblock {Dynamical symmetry of anisotropic singular oscillator}.
\newblock {\em Physics Letters A}, 153:177--180, 1991.

\bibitem{Gao-2013}
S.~Gao, Y.~Wang, and B.~Hou.
\newblock {The classification of Leonard triples of Racah type}.
\newblock {\em Linear algebra and its applications}, 2013.

\bibitem{Gasper-2004}
G.~Gasper and M.~Rahman.
\newblock {\em {Basic Hypergeometric Series}}.
\newblock Cambridge University Press, 2\textsuperscript{nd} edition, 2004.

\bibitem{Zhedanov-2013}
V.X. Genest, L.~Vinet, and A.~Zhedanov.
\newblock {Racah algebra realizations}.
\newblock {\em in preparation}, 2013.

\bibitem{Zhedanov-1988}
Y.~Granovskii and A.~Zhedanov.
\newblock {Nature of the symmetry group of the $6j$-symbol}.
\newblock {\em Soviet Physics JETP}, 67:1982--1985, 1988.

\bibitem{Zhedanov-1989}
Y.~Granovskii and A.~Zhedanov.
\newblock {Exactly solved problems and their quadratic algebras (in Russian)}.
\newblock {\em preprint DonFTI}, 89-7, 1989.

\bibitem{Zhedanov-1991-3}
Y.~Granovskii and A.~Zhedanov.
\newblock {Quadratic algebra as a `hidden' symmetry of the Hartmann potential}.
\newblock {\em Journal of Physics A: Mathematical and General}, 24:3887--3894,
  1991.

\bibitem{Zhedanov-1992-2}
Y.~Granovskii, A.~Zhedanov, and I.~Lutsenko.
\newblock {Mutual integrability, quadratic algebras and dynamical symmetry}.
\newblock {\em Annals of Physics}, 217:1--20, 1992.

\bibitem{Zhedanov-1992}
Y.~Granovskii, A.~Zhedanov, and I.~Lutsenko.
\newblock {Quadratic algebras and dynamics in curved spaces {I}. Oscillator}.
\newblock {\em Theoretical and Mathematical Physics}, 91:474--480, 1992.

\bibitem{Granovskii-1992}
Y.~Granovskii, A.~Zhedanov, and I.~Lutsenko.
\newblock {Quadratic algebras and dynamics in curved spaces {II}. The {K}epler
  problem}.
\newblock {\em Theoretical and Mathematical Physics}, 91:604--612, 1992.

\bibitem{Groenevelt-2005}
W.~Groenevelt.
\newblock {Wilson function transforms related to Racah coefficients }.
\newblock {\em Acta Applicandae Mathematicae}, 91:133--191, 2006.

\bibitem{Koelink-2003}
W.~Groenevelt, E.~Koelink, and H.~Rosengren.
\newblock {\em {Continuous Hahn functions as Clebsch-Gordan coefficients}},
  pages 221--284.
\newblock {Developments in Mathematics}. Springer, 2005.

\bibitem{Koelink-2012}
M.E.H. Ismail and E.~Koelink.
\newblock {Spectral properties of operators using tridiagonalization}.
\newblock {\em Analysis and Applications}, 10, 2012.

\bibitem{Kalnins-1996}
E.G. Kalnins, W.~Miller, and G.~Pogosyan.
\newblock {Superintegrability and associated polynomial solutions: Euclidean
  space and the sphere in two dimensions}.
\newblock {\em Journal of Mathematical Physics}, 37:6439, 1996.

\bibitem{Kalnins-2000}
E.G. Kalnins, W.~Miller, and G.~Pogosyan.
\newblock {Completeness of multiseparable superintegrability in
  $E\under{2,C}$}.
\newblock {\em Journal of Physics A: Mathematical and General}, 33:4105, 2000.

\bibitem{Kalnins-2000-2}
E.G. Kalnins, W.~Miller, and G.~Pogosyan.
\newblock {Completeness of multiseparable superintegrability on the complex
  2-sphere}.
\newblock {\em Journal of Physics A: Mathematical and General}, 33:6791, 2000.

\bibitem{Kalnins-2007}
E.G. Kalnins, W.~Miller, and S.~Post.
\newblock {Wilson polynomials and the generic superintegrable system on the
  2-sphere}.
\newblock {\em Journal of Physics A: Mathematical and Theoretical}, 40:11525,
  2007.

\bibitem{Miller-2011}
E.G. Kalnins, W.~Miller, and S.~Post.
\newblock {Two-Variable Wilson Polynomials and the Generic Superintegrable
  System on the 3-Sphere}.
\newblock {\em SIGMA}, 7:51--76, 2011.

\bibitem{Miller-2013}
E.G. Kalnins, W.~Miller, and S.~Post.
\newblock {Contractions of 2D 2nd order quantum superintegrable systems and the
  Askey scheme for hypergeometric orthogonal polynomials}.
\newblock {\em ArXiv:1212.4766}, 2013.

\bibitem{Koekoek-2010}
R.~Koekoek, P.A. Lesky, and R.F. Swarttouw.
\newblock {\em {Hypergeometric orthogonal polynomials and their
  $q$-analogues}}.
\newblock Springer, 1\textsuperscript{st} edition, 2010.

\bibitem{Kuz-1992}
V.B. Kuznetsov.
\newblock {Quadrics on Riemannian spaces of constant curvature. Separation of
  variables and connection with the Gaudin magnet}.
\newblock {\em Theoretical and Mathematical Physics}, 91:385--404, 1992.

\bibitem{Vinet-1995}
P.~L{\'e}tourneau and L.~Vinet.
\newblock {Superintegrable Systems: Polynomial Algebras and Quasi-Exactly
  Solvable Hamiltonians}.
\newblock {\em Annals of Physics}, 243:144--168, 1995.

\bibitem{Olver-1993}
P.~J. Olver.
\newblock {\em {Applications of {L}ie groups to differential equations}}.
\newblock Springer, 1993.

\bibitem{Winter-2005}
P.~Tempesta, P.~Winternitz, W.~Miller, and G.~Pogosyan, editors.
\newblock {\em {Superintegrability in classical and quantum systems}},
  volume~37.
\newblock AMS, 2005.

\bibitem{Terwilliger-2001}
P.~Terwilliger.
\newblock {Two linear transformations each tridiagonal with respect to an
  eigenbasis of the other}.
\newblock {\em Linear algebra and its applications}, 330:149--203, 2001.

\bibitem{Algebra}
B.~L. van~der Waerden.
\newblock {\em {Algebra}}.
\newblock Spinger-Verlag, 1993.

\bibitem{Klimyk-1991}
N.J. Vilenkin and A.U. Klimyk.
\newblock {\em {Representations of Lie groups and Special Functions}}.
\newblock Kluwer Academic Publishers, 1991.

\bibitem{Winter-1967}
P.~Winternitz, Y.~A. Smorodinsky, M.~Ulh{\`i}r, and I.~Fris.
\newblock {Symmetry groups in classical and quantum mechanics}.
\newblock {\em Soviet Journal of Nuclear Physics}, 4:444--450, 1967.

\bibitem{Zhedanov-1991-1}
A.~Zhedanov.
\newblock {Hidden symmetry of {A}skey-{W}ilson polynomials}.
\newblock {\em Theoretical and Mathematical Physics}, 89:1146--1157, 1991.

\bibitem{Zhedanov-1993}
A.~Zhedanov.
\newblock {Hidden symmetry algebra and overlap coefficients for two ring-shaped
  potentials}.
\newblock {\em Journal of Physics A: Mathematical and General}, 26:4633, 1993.

\end{thebibliography}

\end{document}